\documentclass[aps,twocolumn,superscriptaddress]{revtex4}
\usepackage{graphics}
\usepackage[dvips]{graphicx}
\usepackage{times}
\usepackage{color}
\begin{document}

\title{Defection and extortion as unexpected catalysts of unconditional cooperation in structured populations}

\author{Attila Szolnoki}
\email{szolnoki.attila@ttk.mta.hu}
\affiliation{Institute of Technical Physics and Materials Science, Research Centre for Natural Sciences, Hungarian Academy of Sciences, P.O. Box 49, H-1525 Budapest, Hungary}

\author{Matja{\v z} Perc}
\email{matjaz.perc@uni-mb.si}
\affiliation{Faculty of Natural Sciences and Mathematics, University of Maribor, Koro{\v s}ka cesta 160, SI-2000 Maribor, Slovenia}
\affiliation{CAMTP -- Center for Applied Mathematics and Theoretical Physics, University of Maribor, Krekova 2, SI-2000 Maribor, Slovenia}

\begin{abstract}
We study the evolution of cooperation in the spatial prisoner's dilemma game, where besides unconditional cooperation and defection, tit-for-tat, win-stay-lose-shift and extortion are the five competing strategies. While pairwise imitation fails to sustain unconditional cooperation and extortion regardless of game parametrization, myopic updating gives rise to the coexistence of all five strategies if the temptation to defect is sufficiently large or if the degree distribution of the interaction network is heterogeneous. This counterintuitive evolutionary outcome emerges as a result of an unexpected chain of strategy invasions. Firstly, defectors emerge and coarsen spontaneously among players adopting win-stay-lose-shift. Secondly, extortioners and players adopting tit-for-tat emerge and spread via neutral drift among the emerged defectors. And lastly, among the extortioners, cooperators become viable too. These recurrent evolutionary invasions yield a five-strategy phase that is stable irrespective of the system size and the structure of the interaction network, and they reveal the most unexpected mechanism that stabilizes extortion and cooperation in an evolutionary setting.
\end{abstract}

\maketitle

Widespread cooperation in nature is one of the most important challenges to Darwin's theory of evolution and natural selection, but it is also the main driving force behind the evolutionary transitions that led from single-cell organisms to complex animal and human societies \cite{maynard_95}. And it appears to be this mixture of a fascinating riddle and outmost importance that makes cooperation so irresistibly attractive to study. Evolutionary game theory \cite{maynard_82, weibull_95, hofbauer_98, mestertong_01, nowak_06} is thereby the most frequently employed theoretical framework, revealing mechanisms such as kin selection \cite{hamilton_wd_jtb64a}, network reciprocity \cite{nowak_n92b}, direct and indirect reciprocity \cite{trivers_qrb71, axelrod_s81}, as well as group selection \cite{wilson_ds_an77} as potent promoters of cooperative behavior. Adding to these established five rules for the evolution of cooperation \cite{nowak_s06}, recent years have witnessed a surge of predominantly interdisciplinary studies, linking together knowledge from biology, sociology, economics as well as mathematics and physics, to identify new ways by means of which the successful evolution of cooperation amongst selfish and unrelated individuals can be understood \cite{doebeli_el05, sigmund_tee07, szabo_pr07, roca_plr09, schuster_jbp08, perc_bs10, perc_jrsi13, rand_tcs13}.

From the large array of games that make up evolutionary game theory, none has received as much attention as the prisoner's dilemma game \cite{fudenberg_e86, nowak_n93, santos_prl05, imhof_pnas05, santos_pnas06, tanimoto_pre07, fu_epjb07, gomez-gardenes_prl07, poncela_njp07, fu_pre08b, poncela_epl09, antonioni_pone11, tanimoto_pre12, gracia-lazaro_srep12, gracia-lazaro_pnas12}. Each instance of the game is contested by two players who have to decide simultaneously whether they want to cooperate or defect. The dilemma is given by the fact that although mutual cooperation yields the highest collective payoff, a defector will do better if the opponent decides to cooperate. The rational outcome is thus mutual defection. The popularity of the game was helped significantly by the tournaments that were organized by Robert Axelrod \cite{axelrod_84}, where the most successful strategy for the iterated prisoner's dilemma game was sought. Interestingly the long-term winner was the tit-for-tat strategy by the simple and intuitive virtue of always following the opponent's previous action. However, tit-for-tat cannot correct erroneous moves, and it is also vulnerable to random drift when mutant strategies appear which always cooperate \cite{imhof_jtb07}. Nowak and Sigmund therefore proposed win-stay-lose-shift as another equally simple strategy that has neither of these two disadvantages, and can outperform tit-for-tat in the prisoner's dilemma game \cite{nowak_n93}. Players adopting win-stay-lose-shift simply repeat the previous move if the resulting payoff has met their aspiration level and change otherwise.

The simplicity and effectiveness of strategies like tit-for-tat and win-stay-lose-shift were unrivaled for decades, and they generated a large following of the seminal works that introduced them. Recently, however, Press and Dyson have introduced a new class of so-called zero-determinant strategies that can dominate any opponent in the iterated prisoner's dilemma game \cite{press_pnas12}. A particularly interesting subset of the class are extortion strategies, which ensure that an increase in one's own payoff exceeds the increase in the other player's payoff by a fixed percentage. Extortion is therefore able to dominate any opponent \cite{stewart_pnas12}. But this holds only if players are unable to change strategies in response to their failures. In an evolutionary setting, where players are able to imitate strategies that are more successful, extortion was shown to be evolutionary unstable \cite{adami_ncom13}. If the two players engaged in the game belong to distinct populations, or if the population size is very small, on the other hand, extortioners can nevertheless prevail, and rather counterintuitively, they may also act as catalysts for the evolution of cooperation \cite{hilbe_pnas13}. Evolutionary stability can also be warranted by generous zero-determinant strategies through their mutually supporting behavior \cite{stewart_pnas13}.

Results summarized thus far concerning zero-determinant strategies were obtained in well-mixed populations. Yet it is well-known that stable solutions in structured population can differ significantly from those in well-mixed populations. The most prominent example of this fact is the successful evolution of cooperation in the spatial prisoner's dilemma game through network reciprocity \cite{nowak_n92b}. Further examples include the stabilization of reward \cite{szolnoki_epl10}, peer and pool punishment \cite{helbing_ploscb10, szolnoki_pre11}, in-group favoritism \cite{fu_srep12}, as well as homophily \cite{fu_srep12b}, to name but a few. Indeed, the fact that the interactions among players are frequently not random and best described by a well-mixed model, but rather that they are limited to a set of other players in the population and as such are best described by a network, has far-reaching consequences for the outcome of evolutionary processes \cite{doebeli_el05, szabo_pr07, roca_plr09, perc_bs10, perc_jrsi13}.

Motivated by this, we have recently shown that in structured populations the microscopic dynamic that governs strategy updating plays a decisive role for the fate of extortioners \cite{szolnoki_pre14}. By using the simplest three-strategy model, comprising cooperators ($C$), defectors ($D$), and extortioners ($E_{\chi}$), we have shown that pairwise imitation and birth-death dynamics return the same evolutionary outcomes as reported previously in well-mixed populations. The usage of myopic best response strategy updating, on the other hand, renders extortion evolutionary stable via neutral drift. Counterintuitively, the stability of extortioners helps cooperators to survive even under the most testing conditions, whereby the neutral drift of $E_\chi$ players serves as the entry point, akin to a Trojan horse, for cooperation to grab a hold among defectors. Although the mutually rewarding checkerboard-like coexistence of cooperators and extortioners can always be temporarily disturbed by defectors, it is only a matter of time before the neutral drift reintroduce extortioners and the whole cycle starts anew.

Here we extend our study to five competing strategies, taking into account also the tit-for-tat strategy ($TFT$) and the win-stay-lose-shift strategy ($WSLS$), in addition to the previous three that we have studied in \cite{szolnoki_pre14}. The five strategies $D$, $C$, $E_{\chi}$, $TFT$, and $WSLS$ are the same as studied recently by Hilbe et al.~\cite{hilbe_pnas13} in well-mixed populations, with the strength of the social dilemma $b$ and the strength of exploitation $\chi$ being the two main parameters that determine the payoffs amongst the strategies. For details about the parametrization of the game and the applied updating rules, we refer to the Methods section. The inclusion of the tit-for-tat strategy and the win-stay-lose-shift strategy promises fascinating evolutionary outcomes, especially since under well-mixed conditions $D$ can beat $WSLS$, but the dominance reverses in the presence of the other three strategies. As we will show in the next Section, in structured populations $WSLS$ dominate completely for sufficiently small values of $b$ if the interaction network is characterized by a homogeneous degree distribution. Beyond a threshold value of $b$, or if the interaction network is characterized by a heterogeneous degree distribution (see for example \cite{santos_n08}), however, $D$ emerge and coarsen spontaneously, which in turn opens up the possibility for all the other strategies to emerge as well.

\section*{Results}

\begin{figure}
\centering
\includegraphics[width=8.5cm]{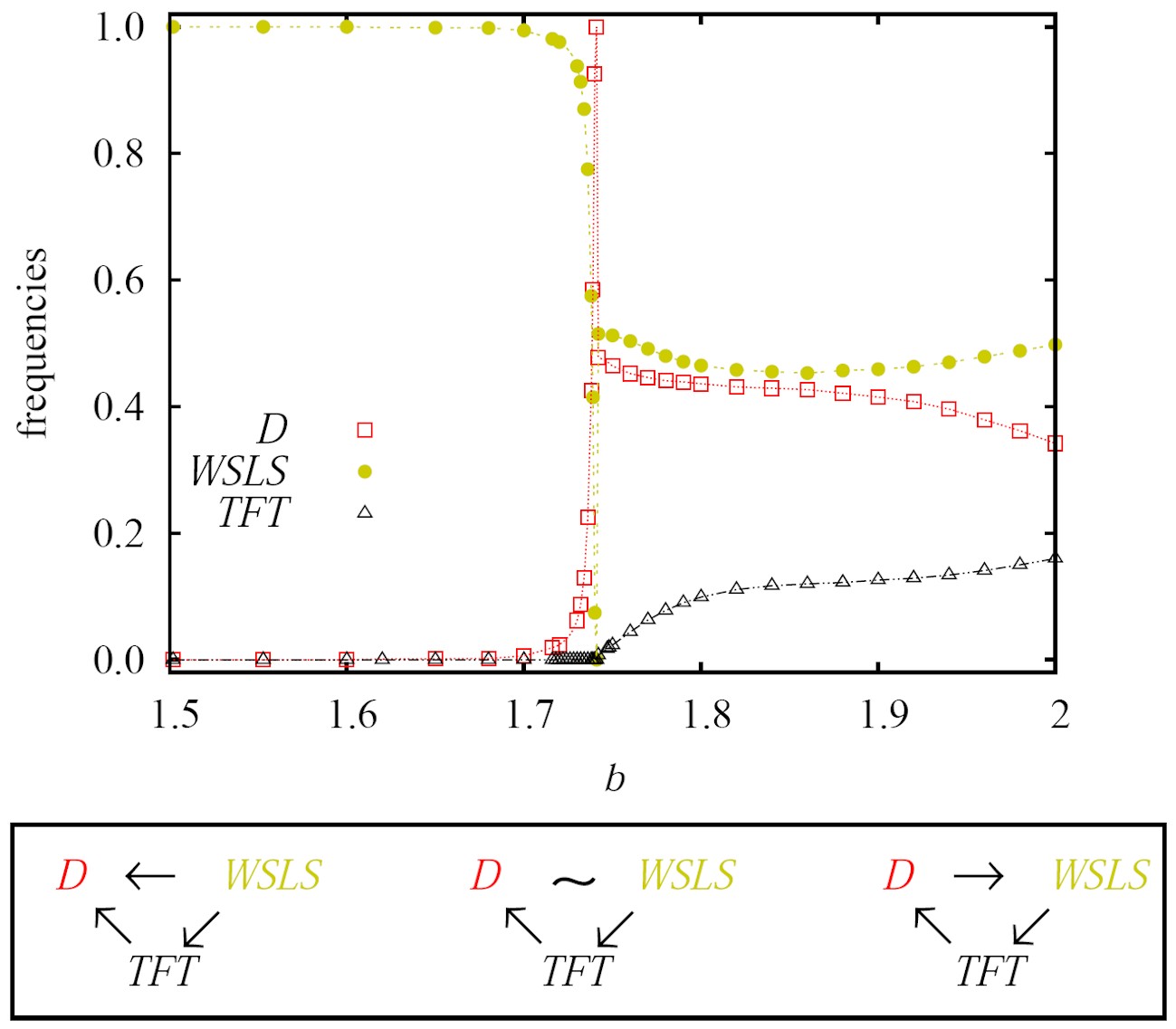}
\caption{Imitation on a square lattice fails to sustain cooperation and extortion. Depicted are the stationary frequencies of surviving strategies in dependence on the strength of the social dilemma $b$. It can be observed that for sufficiently small values of $b$ only $WSLS$ survive. As $b$ increases, the pure $WSLS$ phase first gives way to a narrow two-strategy $WSLS+D$ phase, which then transforms into the three-strategy $WSLS+TFT+D$ phase. The emergence of these three different phases is a direct consequence of dominance relations between the three involved strategies, which are schematically depicted in the bottom frame for the respective values of $b$ from left to right. Arrows show the direction of invasion between strategies.}
\label{imitate}
\end{figure}

Before turning to the main results obtained with myopic best response updating, we present in Fig.~\ref{imitate} the evolutionary outcomes obtained via imitation on a square lattice. If imitation is the basis of strategy updating, then neither cooperators nor extortioners can survive, and this regardless of the strength of the social dilemma and the strength of exploitation. Since extortioners always die out, the composition of the final state is actually completely independent of $\chi$. We have used $\chi=1.5$ for the presented results, but the value influences only the time needed for relaxation towards the final stable solution. Starting with $b \geq 1$ (we show results from $b=1.5$ onwards for clarity with regards to the subsequent phase transitions), the completely dominant strategy is $WSLS$. At the other end of the interval of $b$, we have a stable three-strategy $WSLS+TFT+D$ phase, which is sustained by cyclic dominance. In between, we have a narrow two-strategy $WSLS+D$ phase, which terminates immediately after $D$ reach dominance.

This dependence on $b$ can be understood by considering the relations among the surviving strategies, as summarized in the bottom frame of Fig.~\ref{imitate}. For small values of $b$ (left), $WSLS$ dominate both $D$ and $TFT$. The latter also dominate $D$, but their superior status in this relationship has no effect on the final state. For high values of $b$ (right), the direction of invasion between $WSLS$ and $D$ changes compared to the low $b$ case, while the other two relations remain unchanged. Consequently, instead of a pure $WSLS$ phase, we have a three-strategy $WSLS+TFT+D$ phase, where $WSLS$ invade $TFT$, $TFT$ invade $D$, and $D$ invade $WSLS$ to close the loop of dominance. It is worth emphasizing that this solution is impossible in a well-mixed population for all $b<2$.

\begin{figure}
\centering
\includegraphics[width=8.4cm]{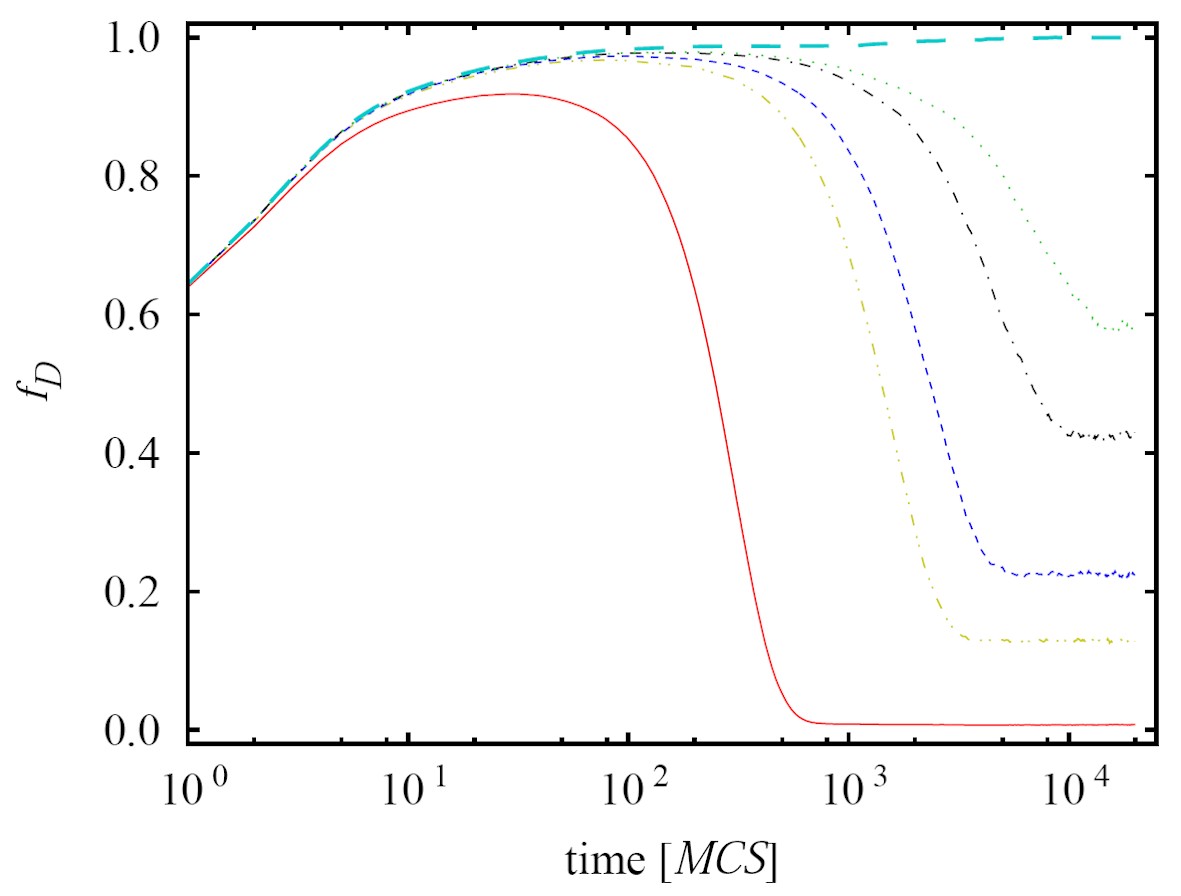}
\caption{The coexistence of defectors and players adopting the win-stay-lose-shift strategy in case of imitation on a square lattice. Depicted is the time evolution of the frequency of defectors $f_D$ as obtained for $b=1.7$, $1.734$, $1.736$, $1.738$, $1.739$ and $1.741$ from bottom to top. The time courses provide insight into the competition for space within the narrow two-strategy $WSLS+D$ phase that can be observed in Fig.~\ref{imitate}. At $b=1.741$ defectors come to dominate the whole population, but their dominance is immediately overthrown in favor of the three-strategy $WSLS+TFT+D$ phase that is sustained by cyclic dominance. The used linear size of the square lattice is $L=1000$. Note that the time scale is logarithmic.}
\label{time}
\end{figure}

In a narrow interval between the pure $WSLS$ phase and the cyclic $WSLS+TFT+D$ phase, we have the situation depicted in the middle of the bottom frame of Fig.~\ref{imitate}, where unlike for small and high values of $b$, the relation between $WSLS$ and $D$ enables their coexistence in a structured population. As for small values of $b$, here too $TFT$ can invade $D$, but this is without effect on the final outcome. The stable two-strategy coexistence is illustrated in Fig.~\ref{time}, where we show how $WSLS$ and $D$ compete for space over time for different values of $b$. The larger the value of $b$, the smaller the fraction of the population that is occupied by $WSLS$ in the stationary state. Interestingly, when $b$ is large enough for $D$ to fully eliminate $WSLS$, the complete dominance of defectors is prevented by the presence of $TFT$, who become viable via a second-order continuous phase transition. From this point onwards, the cyclic dominance $WSLS \to TFT \to D \to WSLS$ starts working until the end of the interval of $b$, as depicted in the main panel of Fig.~\ref{imitate}.

\begin{figure}
\centering
\includegraphics[width=8.5cm]{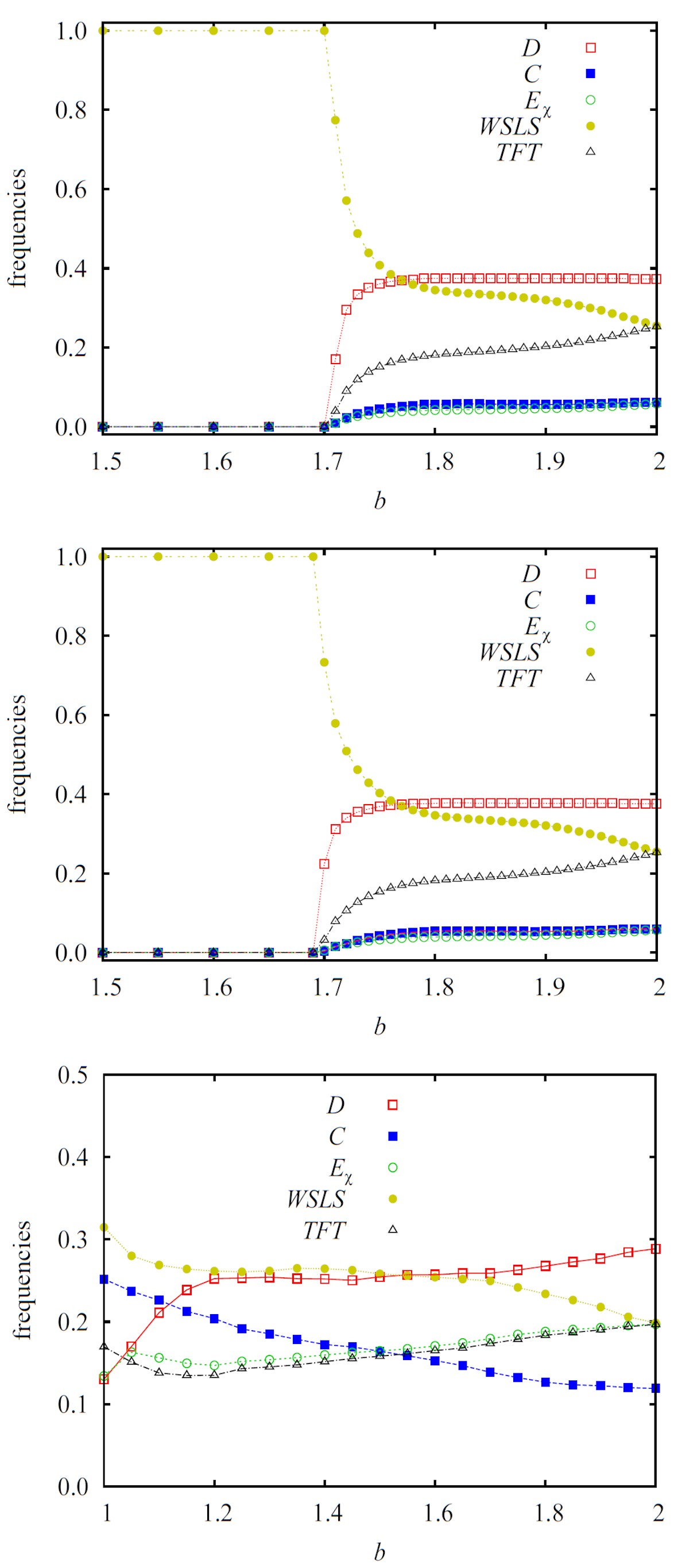}
\caption{Myopic best response updating in structured populations stabilizes extortion and cooperation. Depicted are the stationary frequencies of surviving strategies in dependence on the strength of the social dilemma $b$, as obtained for the strength of extortion $\chi=1.5$ on the square lattice (top), the random regular graph (middle), and the scale-free network (bottom).
It can be observed that players adopting the $WSLS$ strategy dominate for sufficiently small values of $b$ on homogeneous interaction networks (top and middle), but as $b$ increases or if the interaction network is heterogeneous (bottom), the pure $WSLS$ phase gives way to a stable five-strategy $WSLS+D+E_{\chi}+TFT+C$ phase. (\textit{continues on next page})}
\label{myopic}
\end{figure}

Overall, extortion is unable to capitalize on structured interactions if the strategy updating is governed by imitation or a birth-death rule (results not shown), and in fact this is in full qualitative agreement with the results obtained in well-mixed populations \cite{adami_ncom13, hilbe_pnas13}. In the realm of evolutionary games, extortioners do not do well against cooperative strategies like $C$, $TFT$ and $WSLS$. They may thrive for a short period of time, but as soon extortion becomes widespread, it is more profitable to cooperate, which ultimately renders extortion evolutionary unstable.

Myopic strategy updating, on the other hand, can sustain very different evolutionary outcomes as it allows players to adopt strategies that are not necessarily present in their interaction neighborhood. In fact, strategies need not be present in the population at all, as long as they are an option for the players to choose randomly when it is their turn to perhaps change their strategy. Nevertheless, we emphasize that myopic best response updating is different from mutation, because each individual strategy change is still driven by the payoff difference, as described by Eq.~\ref{myop}. Results presented in Fig.~\ref{myopic} obtained on the square lattice (top) and the random regular graph (middle) show that for sufficiently small values of $b$ the final state is the same as under imitation dynamics. Players adopting $WSLS$ dominate completely from $b=1$ onwards (as in Fig.~\ref{imitate}, we show results for $b \geq 1.5$ only). At a critical value of $b$, however, a second-order continuous phase transition rather unexpectedly leads to the stable coexistence of all five competing strategies. A similar diversity of strategies prevails on heterogeneous interaction networks, as illustrated by the results obtained on a scale-free network shown in the bottom panel of Fig.~\ref{myopic}. Myopic best response updating is thus able to stabilize extortion in structured populations. Perhaps even more surprisingly, as the strength of the social dilemma increases, the two cooperative strategies $C$ and $TFT$ become viable as well. This outcome is rather independent of the structure of the interaction network.

\begin{figure}\renewcommand{\thefigure}{3}
\caption{(\textit{continues from previous page}) Here defectors emerge and coarsen spontaneously because for sufficiently large values of $b$ their payoff becomes larger than that of clustered $WSLS$ players. The emergence of defectors immediately opens the door to the survival of extortioners and $TFT$ players, which both emerge by chance and spread by means of neutral drift. Lastly, with the emergence of extortioners and $TFT$ players cooperators become viable as well, thus forming the stable five-strategy phase. The latter is virtually unaffected by different values of $\chi$, as demonstrated in Fig.~\ref{chi}. Importantly, the described coexistence of the competing strategies is a universal behavior that can be observed in structured populations regardless of the properties of the interaction network, and even across the whole span of $b$ values, as illustrated in the bottom panel. Characteristic snapshots depicting the described key stages of the evolutionary process are presented in Fig.~\ref{snaps}.}
\end{figure}

\begin{figure}\renewcommand{\thefigure}{4}
\centering
\includegraphics[width=8.5cm]{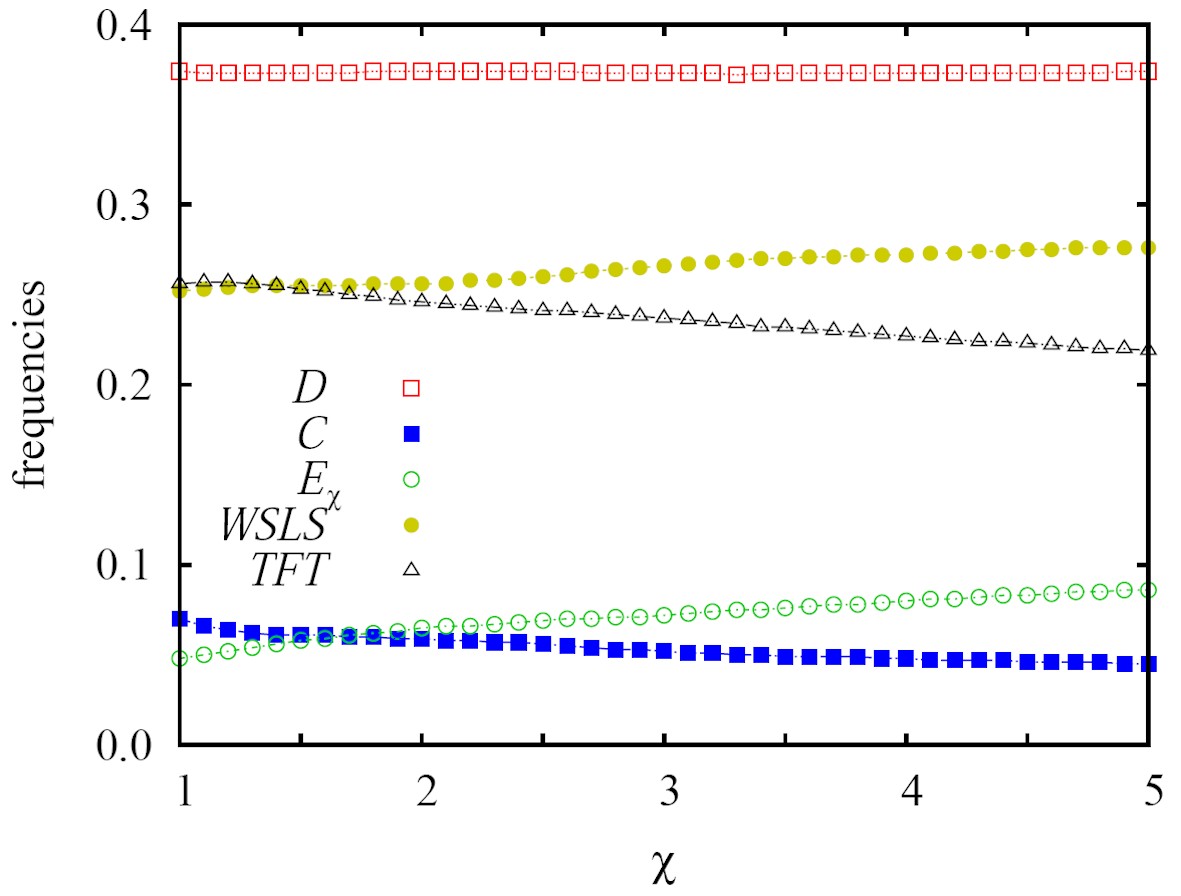}
\caption{The strength of extortion has a negligible impact on the stationary frequencies of competing strategies, and it does not affect the evolutionary stability of extortion and cooperation. Depicted are the stationary frequencies of surviving strategies in dependence on the strength of extortion $\chi$, as obtained for the social dilemma strength $b=2$ on a square lattice. It can be observed that the variations of all frequencies are small. Expectedly, larger values of $\chi$ favor extortion. The neutral drift of $TFT$ players therefore becomes slightly less prolific, which in turn also slightly decreases the frequency of cooperators. Interestingly, the stationary frequencies of strategies at $b=2$ and their $\chi$-dependency are practically indistinguishable for the square lattice and the random regular graph. This further highlights the irrelevance of the structure of the interaction network under myopic best response updating, and thus also the  universality of the presented results.}
\label{chi}
\end{figure}

Since extortioners survive for sufficiently high values of $b$, the strength of extortion $\chi$ might play a role too, but as evidenced by the results presented in Fig.~\ref{chi}, this role is in fact very minor. As the value of $\chi$ increases, the extortioners become slightly more common on the expense of $TFT$ and $C$ players, but overall this does not affect the evolutionary stability of extortion and cooperation. Compared to our previous results presented in \cite{szolnoki_pre14}, where we have studied the three strategy variant of the game without $TFT$ and $WSLS$ players, the role of $\chi$ is less significant here mainly because the stationary frequency of extortioners is much smaller. The fact that their frequency is much smaller, however, is a direct consequence of the presence of the two additional cooperative strategies ($TFT$ and $WSLS$), which in turn highlights the general subordinate role of extortioners compared to cooperation in evolutionary games. The latter was emphasized already in \cite{adami_ncom13, hilbe_pnas13}, as well as by the results presented in Fig.~\ref{imitate} above. Also contributing to the minor role of $\chi$ is that the emergence of extortioners is in fact a second-order effect, as we will explain next.

\begin{figure}\renewcommand{\thefigure}{5}
\centering
\includegraphics[width=8.5cm]{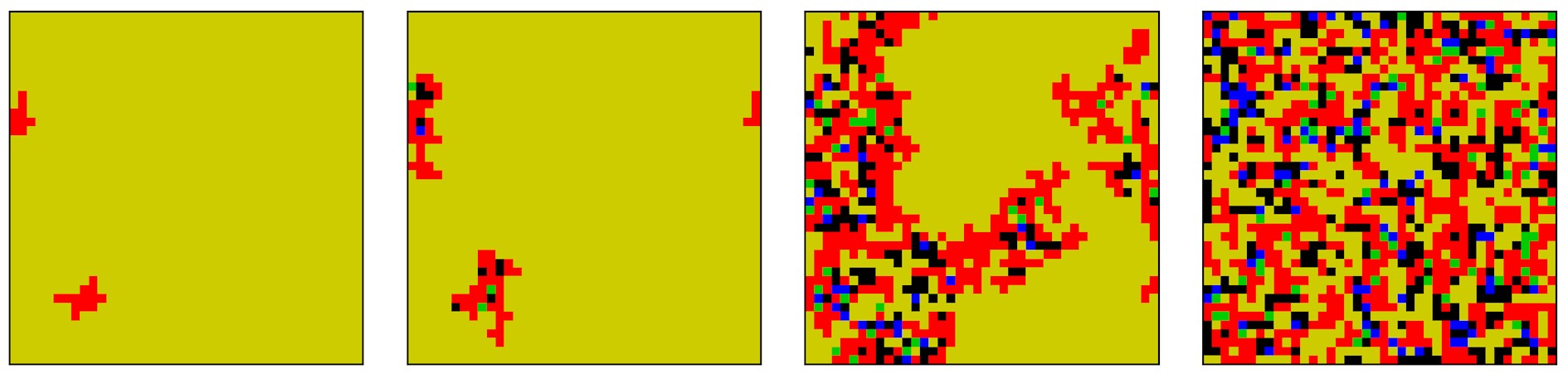}
\caption{Characteristic time evolution of the spatial distribution of the five competing strategies on a square lattice. The evolution starts from the full $WSLS$ phase (not shown), using $b=1.8$ and $\chi=1.5$. At $MCS=5$ (leftmost panel), first defectors start emerging because their payoff is comparable with $WSLS$ players. Soon thereafter, at $MCS=10$ (second panel from left), first extortioners and $TFT$ players emerge. Both have neutral relations with the defectors, and thus their emergence and spreading are due to chance and neutral drift. At $MCS=30$, as soon as locally the number of extortioners becomes sufficiently large, cooperators emerge as well due to their higher payoffs, and their spreading is additionally supported by the $TFT$ players. The recurrence of these elementary processes eventually spreads a stable mixture of all five strategies across the whole population, as depicted in the rightmost panel that was taken at $MCS=100$. The color encoding of the strategies is the same as used in Figs.~\ref{myopic} and \ref{chi}. For clarity with regards to individual players and their strategies, we have used a small square lattice with linear size $L=40$.}
\label{snaps}
\end{figure}

To understand why $E_{\chi}$, $TFT$ and $C$ emerge as $b$ increases, it is instructive to consider the erosion of the pure $WSLS$ phase on square lattice, as illustrated in Fig.~\ref{snaps}. For a sufficiently high value of $b$ defectors emerge and start coarsen spontaneously because their payoff becomes competitive with the payoff of aggregated $WSLS$ players. The emergence of the $D$ phase, however, paves the way for the emergence of all the other strategies. Namely, both $E_{\chi}$ and $TFT$ are neutral against $D$, and thus they may emerge by chance and spread via neutral drift. As $E_{\chi}$ accumulate locally, $C$ become viable too because their payoff is higher. The emergence of $C$ is helped further (or at least not hindered) by $TFT$, who are neutral with $C$. During this unexpected chain of strategy invasions, defection and extortion thus emerge as catalysts of unconditional cooperation. Effectively, the defectors act as a Trojan horse for all the other strategies, while subsequently the extortioners act as a Trojan horse for cooperation. Evidently, the spreading of $C$, which utilizes the neutral drift of $E_\chi$, will be controlled by defectors and $WSLS$ players who can strike back since their presence in place of an extortioner may yield a higher payoff in a predominantly cooperative neighborhood. This, however, will again be only temporary, since the described elementary invasions are bound to recur, thus assuring the stability of the $WSLS+D+E_{\chi}+TFT+C$ phase.

An important lesson learned from the presented results in Fig.~\ref{snaps} is that although extortion can be as counterproductive as defection, it is still less destructive. For an unconditional cooperator it never pays sticking with the strategy if surrounded by defectors, but it may be the best option among extortioners. Cooperators are of course happiest among other cooperators, but in the presence of extortioners they can still attain a positive payoff, and this is much better than nothing or a negative value in the presence of defectors. It is worth emphasizing that this argument is valid independently of the properties of the interaction network, as the described chain of strategy invasions emerges in all the structured populations that we have considered.

\section*{Discussion}
We have shown that even if the set of competing strategies is extended to encompass, besides unconditional cooperators, defectors and extortioners \cite{szolnoki_pre14}, also the tit-for-tat strategy and the win-stay-lose-shift strategy, the imitation dynamics in structured populations is still unable to render extortion evolutionary stable. For sufficiently small values of $b$ only players adopting the win-stay-lose-shift strategy survive, while beyond a threshold value a stable three-strategy phase consisting of defectors, tit-for-tat and win-stay-lose-shift players emerges. Since extortioners never survive, the strength of exploitation $\chi$ is without effect. These results agree with those reported previously for sizable isolated well-mixed populations \cite{hilbe_pnas13}, and they highlight the severe challenges that extortioners face when vying for survival in the realm of evolutionary games where players are able to imitate strategies that are performing better \cite{adami_ncom13}.

If the evolution is governed by myopic best response updating, however, the outcomes are significantly different from those obtained via imitation. We have shown that for sufficiently large values of $b$ the complete dominance of win-stay-lose-shift players is broken as soon as defectors emerge and start coarsening. Subsequently, within the homogeneous domains of defectors, extortion becomes viable too via the same mechanism as we have described before in \cite{szolnoki_pre14}. In particular, extortioners and defectors are neutral, and hence the former can emerge by chance and spread via neutral drift. Yet as soon as extortioners emerge, cooperators can finally emerge as well, because in competition with the former they are superior. In this evolutionary scenario, defection and extortion thus act as the most surprising catalysts of unconditional cooperation in structured populations. Moreover, we have shown that the coexistence of all competing strategies occurs across the whole interval of $b$ values if a heterogeneous (scale-free) network describes the interactions among players. Because of this unlikely path towards cooperation, we conclude that defectors and extortioners effectively play the role of a Trojan horse for cooperators. Interestingly, similar transient roles of extortionate behavior were recently reported in the realm of well-mixed populations when studying the adaptive dynamics of extortion and compliance \cite{hilbe_pone13b}. Moreover, after the emergence and coarsening of defectors, in the presently studied game the tit-for-tat players also become viable as they are likewise neutral, and can thus spread via neutral drift just like extortioners. In recurrence, these evolutionary processes give rise to a stable five-strategy phase that is hardly affected by the strength of exploitation $\chi$, and it is also robust to the population size and the structure of the interaction network.

Taken together, these results thus have a high degree of universality and highlight the relevance of coarsening, the emergence of role-separating strategy distributions (which manifests as checkerboard ordering on regular graphs), and best response updating in evolutionary games. The latter is especially important, as it appears to be an integral part of human behavior \cite{matsui_jet92, blume_l_geb93, ellison_econm93}. From the more pragmatical point of view, best response updating conveys to the players an ability to explore the space of available strategies even if they are not present in their immediate neighborhood or even in the population as a whole, and by doing so, such updating dynamics opens up the door to the most counterintuitive evolutionary outcomes. Similarly to kin competition, the presented results also highlight the other side of network reciprocity. Namely, it does not only support cooperative behavior by means of clustering, but it also reveals the consequences of bad decisions -- defectors and extortioners become weak when they become surrounded by their like. From this point of view, it is understandable and indeed expected that structured populations, if anything, hinder the successful evolution of extortion under imitation. The surprising positive role of extortioners becomes apparent only under best response updating, where the threatening loom of widespread defection is drifted away by the lesser evil to eventually introduce more constructive cooperative strategies.

\section*{Methods}
We adopt the same game parametrization as Hilbe et al. \cite{hilbe_pnas13}. Accordingly, the payoff matrix for the five competing strategies is\\ \\
\centerline{\begin{tabular}{r|c c c c c}
 & $TFT$ & $WSLS$ & $E_\chi$ & $all \,C$ & $all \,D$ \\
\hline
$TFT$ & $\frac{1}{2}$ &$\frac{1}{2}$ &$0$ &$1$ &$0$\\
$WSLS$ & $\frac{1}{2}$ &$1$ &$\frac{(2b-1)\chi}{3b-2+(3b-1)\chi}$ &$\frac{b+1}{2}$ &$\frac{1-b}{2}$\\
$E_\chi$ & $0$ &$\frac{(2b-1)\chi}{3b-2+(3b-1)\chi}$ &$0$ &$\frac{(2b-1)\chi}{b-1+b\chi}$ &0 \\
$all \,C$ & $1$ &$\frac{2-b}{2}$ &$\frac{2b-1}{b-1+b\chi}$ &$1$ &$1-b$ \\
$all \,D$ & $0$ &$\frac{b}{2}$ &$0$ &$b$ &$0$ \\
\end{tabular}}\\ \\ \\
where $b$ is the benefit to the other player provided by each cooperator at the cost $c$, and $\chi$ determines the surplus of the extortioner in relation to the surplus of the other player. Moreover, we use $b-c=1$, thus having $b>1$ and $\chi>1$ as the two main parameters. The former determines the strength of the social dilemma, while the latter determines just how strongly strategy $E_\chi$ exploits cooperators. A direct comparison of the extortioner strategy with the other strategies reveals that $E_\chi$ is neutral with unconditional defectors and players adopting the $TFT$ strategy. The latter, however, may beat $E_\chi$ if they are surrounded by other $TFT$ players. Similar relations hold for the competition between $E_\chi$ and $WSLS$ players. While the latter receive the same income from a direct interaction, they do gain more if the neighbors also adopt the $WSLS$ strategy. It is also worth noting that the payoffs between $C$ and $D$ constitute the so-called donation game, which is an important special case of the iterated prisoner's dilemma game with all the original properties retained \cite{brede_pone13}.

We predominantly consider a $L \times L$ square lattice with periodic boundary conditions as the simplest interaction network to describe a structured population. To demonstrate the robustness of our findings, we also use a random regular graph and the scale-free network with the same average degree, which is likely somewhat more apt to describe realistic social and technological networks \cite{barabasi_s99}. We have used population sizes from $10^4$ up to $10^6$ players to avoid finite-size effects.

Unless stated differently, for example to illustrate a specific invasion process as in Fig.~\ref{snaps}, we use random initial conditions such that all five strategies are uniformly distributed across the network. We carry out Monte Carlo simulations comprising the following elementary steps. First, a randomly selected player $x$ with strategy $s_x$ acquires its payoff $p_x$ by playing the game with its $k$ neighbors, as specified by the underlying interaction network. Next, player $x$ changes its strategy $s_x$ to $s_x^{\prime}$ with the probability
\begin{equation}
q(s_x^{\prime} \to s_x) =\frac{1}{1+\exp[(p_x-p_x^{\prime})/K]}\ \,
\label{myop}
\end{equation}
where $p_x^{\prime}$ is the payoff of the same player if adopting strategy $s_x^{\prime}$ within the same neighborhood, and $K=0.05$ quantifies a small uncertainty that is related to the strategy adoption process \cite{szabo_pr07}. The strategy $s_x^{\prime}$ should of course be different from $s_x$, and it is drawn randomly from the remaining four strategies. Such strategy updating is known as the myopic best response rule \cite{matsui_jet92}.

We also consider the more traditional strategy imitation, where player $x$ imitates the strategy of a randomly selected neighbor $y$, only that $p_x^{\prime}$ in Eq.~\ref{myop} is replaced by $p_y$ \cite{szabo_pr07}, as well as death-birth updating as described for example in \cite{ohtsuki_jtb06}. Regardless of the applied strategy updating rule, we let the system evolve towards the stationary state where the average frequency of strategies becomes time independent. We measure time in full Monte Carlo steps ($MCS$), during which each player is given a chance to change its strategy once on average.

\begin{acknowledgments}
This research was supported by the Hungarian National Research Fund (Grant K-101490), TAMOP-4.2.2.A-11/1/KONV-2012-0051, and the Slovenian Research Agency (Grants J1-4055 and P5-0027).
\end{acknowledgments}

\end{document}